\documentstyle[aps,prl,epsfig]{revtex}

\begin{document}

\title{Hadron Formation in Deep-Inelastic Positron Scattering in a
Nuclear Environment}

\vspace{5 mm}

\author{
\centerline {\it The HERMES Collaboration}\medskip \\
\ \\ 
A.~Airapetian$^{30}$,
N.~Akopov$^{30}$,
Z.~Akopov$^{30}$,
I.~Akushevich$^{7}$,
M.~Amarian$^{23,25,30}$,
J.~Arrington$^{2}$,
E.C.~Aschenauer$^{7,13,23}$,
H.~Avakian$^{11}$,
R.~Avakian$^{30}$,
A.~Avetissian$^{30}$
% ...
} 

\address{
% ...
$^2$Physics Division, Argonne National Laboratory, Argonne, Illinois 60439-4843, USA\\
% ...
$^7$DESY Zeuthen, 15738 Zeuthen, Germany\\
% ...
$^{11}$Istituto Nazionale di Fisica Nucleare, Laboratori Nazionali di Frascati, 00044 Frascati, Italy\\
% ...
$^{13}$Department of Subatomic and Radiation Physics, University of Gent, 9000 Gent, Belgium\\
% ...
$^{23}$Nationaal Instituut voor Kernfysica en Hoge-Energiefysica (NIKHEF), 1009 DB Amsterdam, The Netherlands\\
% ...
$^{25}$Istituto Nazionale di Fisica Nucleare, Sezione Roma 1 --
Gruppo Sanit\`a and Physics Laboratory, Istituto Superiore di Sanit\`a, 00161 Roma, Italy\\
% ...
$^{30}$Yerevan Physics Institute, 375036, Yerevan, Armenia\\
% ...
}
 
\date{\today}
\maketitle

\begin{abstract}
The influence of the nuclear medium on the production of charged hadrons
in semi-inclusive deep-inelastic scattering has been studied by the 
HERMES experiment at DESY using a 27.5 GeV positron beam.
The differential multiplicity of charged hadrons and identified
charged pions from nitrogen relative 
to that from deuterium has been measured as a function of 
the virtual photon energy $\nu$ and the fraction $z$ of this energy 
transferred to the hadron. There are observed substantial reductions 
of the multiplicity ratio $R_M^{h}$ 
at low $\nu$ and at high $z$, both of which are well described by 
a gluon-bremsstrahlung
model of hadronization. A significant difference of the 
$\nu$-dependence of $R_M^{h}$ is found between positive and negative hadrons. 
This is interpreted in terms of a difference
between the formation times of protons and pions,
using a phenomenological model to describe 
the $\nu$- and $z$-dependence of $R_M^{h}$.
\end{abstract}

\medskip

\vspace{0.5cm}

\centerline{PACS numbers: 13.60.Hb, 13.60.Le, 13.60.-r, 24.85.+p}

\twocolumn

In deep-inelastic scattering (DIS) an incident lepton interacts with
a target via the exchange of a gauge boson between the lepton
and a parton (a quark or an antiquark). The struck parton is
subjected to a sequence of hard parton- and soft hadron-production processes,
resulting in the formation of hadrons.
By carrying out DIS experiments on nuclear targets,
it is possible to study the hadronization
process during the time period immediately after the 
quark has been struck by the virtual photon~\cite{Bial89}. 
In the simplest scenario the nucleus, which has the size
of a few fm, acts as an ensemble of targets with which
the struck quark or the produced hadron may interact.
If an interaction occurs, the number of leading
hadrons produced per DIS event and per nucleon  
is reduced compared to that for a free nucleon. 
The reduction of hadron multiplicity
depends on the distance traversed by the struck
quark before the hadron is formed, the (unknown) quark-nucleon cross
section and the (known) hadron-nucleon cross section.
Hence, measurements  of the multiplicity of hadrons produced
on nuclei can provide information on the space-time
structure of the hadronization process.

In this paper we present the results of deep-inelastic positron scattering
measurements on $^{14}$N and $^{2}$H targets
carried out by the HERMES Collaboration at DESY.
The reduction of the hadron multiplicity on $^{14}$N relative to 
that on $^{2}$H has been determined from the data.
Deuterium was used as a reference target instead of hydrogen to account 
for the difference in scattering from neutrons or protons.

In the past, semi-inclusive leptoproduction of hadrons from nuclei was studied 
at SLAC with electrons~\cite{osborne} and
at CERN and FNAL with high-energy muons by EMC~\cite{EMC} and
E665~\cite{Xe}. In these experiments the effect of the nucleus 
on the early stages of hadronization appeared in the dependence
of the multiplicity ratios on the energy $\nu$ of the virtual 
photon (as defined in the target rest frame).  
The effect is most prominent if $\nu$ ranges from a few 
GeV to a few tens of GeV. The $\nu$-range covered by the present 
experiment (7 -- 23 GeV was used in the present analysis) 
is low enough to be optimal for studying the influence of the nuclear 
medium on the hadronization process, while it is still high enough
to be in the scaling regime, i.e. for the DIS-framework to apply.

The theoretical description of hadronization is difficult,
as it is impossible to treat the problem in
a perturbative QCD-framework. Although the
nuclear environment can be used to study certain aspects of 
the hadron formation process, the same difficulties
apply to the description of nuclear effects
on hadronization. Therefore, one has to resort to model calculations.
The HERMES data on hadron production from nuclei are compared to the
gluon-bremsstrahlung-model calculations of Ref.~\cite{boris},
and the phenomenological intranuclear reinteraction model of
Refs.~\cite{1timescale,2timescale}. 

In the gluon-bremsstrahlung model, the struck quark
is assumed to lose 
energy via the emission of gluons until, in the case of 
meson formation, a $q\bar{q}$ configuration
is formed consisting of the struck quark and an antiquark originating
from the last emitted gluon~\cite{boris,kop90,kop85,brodhoy}.
If this last step occurs inside the nucleus, the nuclear environment 
affects the hadron multiplicity because the
meson interacts with the nuclear medium
with a sizable cross section.
Moreover, the interaction of the initial
quark with the nuclear medium causes the emission of additional
soft gluons. On the other hand, 
the initial --- possibly small ---  $q\bar{q}$ configuration
represents a color dipole which
may have a reduced probability of interaction with its environment
(Color Transparency~\cite{brod,ct}).
An estimate of the combined effect of the soft gluon radiation and
Color Transparency on the multiplicity ratio in the framework
of the gluon-bremsstrahlung model~\cite{boris}
yields 2--3\% for the kinematics of the present experiment.

An important parameter characterizing the hadronization process 
is the formation time (in the target rest frame) $t_f^h$, 
which is defined as the mean time elapsed 
between the moment the quark is struck and the creation 
of the leading hadron $h$. In the gluon-bremsstrahlung model~\cite{boris},
the probability distribution of $t_f^h$ may be evaluated for given values of
$z$ and $\nu$, where
$z$ represents the fraction of the energy $\nu$ transferred to the hadron.
The mean value of $t_f^h$ exhibits an approximate
$(1-z)\nu$-dependence.
In the phenomenological models~\cite{1timescale,2timescale},
on the other hand, the formation time is assumed
to be equal to the product of a time constant $\tau_h$
intrinsic to the kind of hadron produced,
and a time-dilation factor $\nu/m$ where
$m$ represents a mass. 
In the past, several functional forms for $t_f^h$ have been 
considered~\cite{1timescale,2timescale,brod,chmaj,lowgott,pavel} --- for 
instance $t_f^h = \tau_h  z\nu/m_h$ or $t_f^h = \tau_h \nu/m_q$, where
$m_h$ ($m_q$) represents the mass of the hadron (constituent quark). 
Alternatively, one may take for $m$ the kinematic
expression derived by Berger~\cite{berger} for the
mass of the offshell struck (current) quark. Under the assumption that
the mass of the initial quark and that
of the produced hadron can be neglected, one finds
$t_f^h = \tau_h \sqrt{z(1-z)/p_T^{2}} \nu$, where $p_T$
represents the component of the hadron momentum transverse
to the direction of the virtual-photon momentum.

The experimental results are presented in terms of the 
multiplicity ratio $R_M^{h}$, which 
represents the ratio of the number of hadrons of type $h$
produced per DIS event for 
a nuclear target of mass A to that from a deuterium target (D):
\begin{equation}
R_M^{h}(z,\nu) = { { \frac{N_h^A(z,\nu)}{N_e^A(\nu)} }  \Biggm/
                   { \frac{N_h^D(z,\nu)}{N_e^D(\nu)} }  }
\label{eq:att}
\end{equation}
\noindent
with $N_h(z,\nu)$ the number of semi-inclusive hadrons in a 
given ($z,\nu$)-bin, and
$N_e(\nu)$ the number of inclusive DIS positrons in the same $\nu$-bin.
For the purpose of the present analysis, the multiplicity ratio 
was determined as a function of $\nu$ and $z$, while
integrating over all other kinematic variables. 
The data for $R_M^{h}$ are only weakly dependent on
either $Q^2$ or $p_T^2$~\cite{hunen}.

The HERMES experiment at DESY makes use of
the 27.5 GeV positron storage ring of HERA.
The $^{14}$N and $^2$H target gases are injected into a tubular open-ended
storage cell through which the beam passes.
The cell provides a 40 cm long target with areal densities
of up to 6 $\times$ 10$^{15}$ nucleons/cm$^2$ for $^{14}$N.
The dead time of the data acquisition system was observed
to be less than 5\% even at the highest
luminosities of about 10$^{33}$ cm$^{-2}$s$^{-1}$.

In the HERMES spectrometer~\cite{hermesdetector} both the 
scattered positrons 
and the produced hadrons are detected and identified within an
angular acceptance of $\pm$ 170 mrad horizontally, and $\pm$ 
(40 -- 140) mrad vertically. The trigger for scattered positrons 
was formed from a coincidence between 

\begin{figure} [th]
\begin{center}
\includegraphics[width=\hsize]{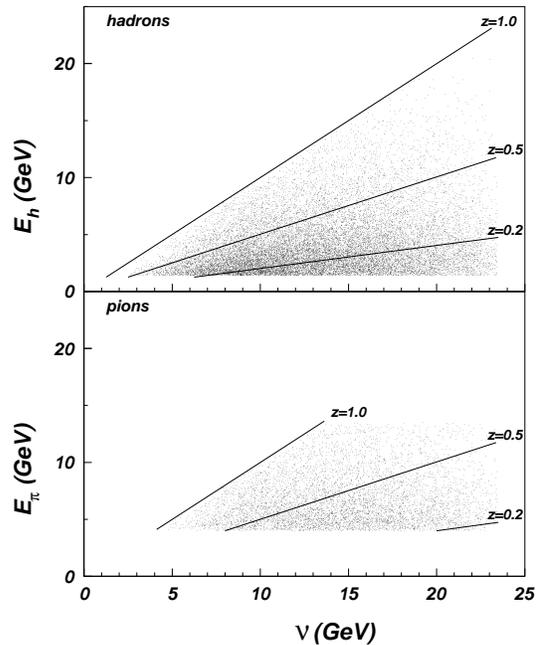}
\end{center}
\nopagebreak
\vspace{-0.5cm}
\caption[]{\label{fig:correl}
Scatter plot of the hadron (or pion)
energy $E_h$ ($E_\pi$) and the energy transfer $\nu$. Lines
representing constant values of $z$ are shown as well.}
\end{figure}

\noindent
several scintillator
hodoscope planes and a lead-glass calorimeter.
The trigger required an energy of more than 3.5 GeV deposited
locally in the calorimeter.
Positron identification was accomplished using
the calorimeter, the second hodoscope, which functioned as a
preshower counter, a transition-radiation detector,
and a threshold gas \v Cerenkov counter filled with a mixture
of $N_2$ and $C_4F_{10}$
at atmospheric pressure. This system provided
positron identification with an average efficiency of 99~\% and a
hadron contamination of less than 1~\%. Hadrons with an energy
$E_h \ge$ 1.4 GeV could be distinguished from leptons.
Furthermore, in the momentum range between 4 and 13.5 GeV,
pions were identified with 
the help of the \v Cerenkov counter.

Events were selected by imposing constraints on 
the four-momentum squared $-Q^2$ of the virtual photon,
the invariant 
mass of the photon-nucleon system $W$, the Bjorken scaling variable 
$x=\frac{Q^2}{2m_p \nu}$ (with $m_p$ the proton mass), and the energy 
fraction of the
virtual photon $y=\nu/E_{beam}$. For each event it was required that $x>0.06$,
$Q^2>1$~GeV$^2$, $W >2$~GeV, and $y<$~0.85.
The requirements on $W$ and $y$ are applied to exclude nucleon resonances 
and to limit the magnitude of the radiative corrections, respectively.
The constraints on ($x,Q^2$) are applied to exclude the 
kinematic region where an anomalous ratio of the longitudinal to 
transverse cross sections for inclusive deep-inelastic
scattering from $^{14}$N has been observed~\cite{f2ratio}.
The anomaly can be interpreted as being due to the absorption of the 
virtual photon by a correlated quark pair~\cite{barshay} or by
a nuclear meson~\cite{MBK}. Either interpretation breaches the
assumption of incoherent
lepton-quark scattering, inherent in deep-inelastic scattering.

The instrumental threshold on $E_h$ is larger for pions than for
hadrons. This implies, as is illustrated in Fig.~\ref{fig:correl},
that e.g. the pion $z$-acceptance is restricted to rather large
values as $\nu$ decreases. Hence, to ensure that the $\nu$-dependence
of the multiplicity ratio
does not correspond to a strong variation of the mean $z$-value, the 
presented data are confined to
$z>$ 0.2 and $\nu>$ 7 GeV for hadrons and
$z>$ 0.5 and $\nu>$ 8 GeV for pions.

Under the kinematic constraints described above, the number of
selected DIS events is 0.88 (1.05) $\times 10^6$ for 
$^{14}$N ($^{2}$H) if $\nu >$ 7 GeV. For $\nu >$ 8 GeV
these numbers decrease to 0.76 $\times 10^6$ for $^{14}$N and 
0.91 $\times 10^6$ for $^{2}$H.
The number of hadrons with $z$ $>$ 0.2 equals
227 (288) $\times 10^3$, and the number of pions with $z$ $>$ 0.5
is 36 (48) $\times 10^3$ for $^{14}$N ($^{2}$H).

The HERMES data have been corrected for radiative processes
using the codes of Refs.~\cite{Tera,Akush}. As is commonly
done~\cite{osborne,EMC},
the contributions from nuclear elastic scattering, quasi-elastic
scattering and DIS were treated independently. Coherence-length
effects~\cite{Kopel} are not included in this approach.
The code of Ref.~\cite{Akush}
was modified to include the measured semi-inclusive DIS cross sections for 
$^{14}$N. The size of the radiative correction was found to be negligible
in most of the kinematic range, with a maximum of 3\% at the highest
value of $\nu$. The correction is so small because most
of the contributions to the radiative corrections
cancel in the multiplicity ratio.

The systematic uncertainty of the present data is 3\% or less. The
uncertainty is small due to the fact that double ratios of 
semi-inclusive and inclusive yields are measured. The main contributions

\begin{figure} [th]
\begin{center}
\includegraphics[width=\hsize]{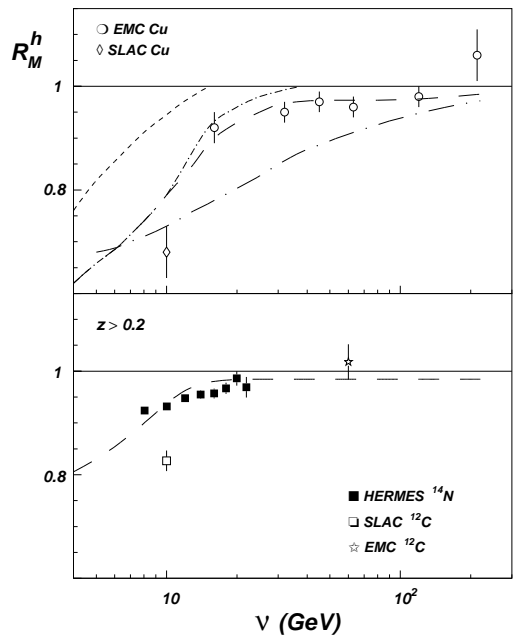}
\end{center}
\nopagebreak
\vspace{-0.5cm}
\caption[]{\label{fig:world}
Charged hadron multiplicity ratio $R_M^{h}$
as a function of $\nu$ for values of $z$ larger
than 0.2. In the upper panel the CERN~\cite{EMC} and SLAC~\cite{osborne}
data for Cu are compared to various phenomenological calculations taken
from the original publications~\cite{EMC,pavel}.
In the lower panel the HERMES data on $^{14}$N are represented by
solid squares, while the open star represents the CERN data point on
$^{12}$C and the open square the SLAC data point on $^{12}$C. The error bars
represent the statistical uncertainty only. The systematic
uncertainty of the HERMES data is $\le$ 3\%. The curves are
described in the text.}
\end{figure}

\noindent
to the systematic uncertainty arise from radiative corrections 
($<$ 2\%), overall efficiency ($<$ 1.5\%),
and diffractive $\rho^0$-meson production
($<$1.5\%)~\cite{hunen}.
It has been verified that the acceptance
for semi-inclusive hadron production is the same for both targets,
by plotting the multiplicity ratio versus
hadron angle and $p_T^2$. 
These ratios are constant to within the expected statistical precision
except for the data beyond $p_T^2$ = 1 GeV$^2$, where an enhancement
similar to that 
reported by Ref.~\cite{EMC} is observed. Because
of the steep decline of semi-inclusive hadron production
with $p_T^2$,
this enhancement does not contribute to the present data.  
The nuclear attenuation of pions resulting from the decay of
$\rho$-mesons
formed in the fragmentation process (which constitute about
20\% of the semi-inclusive pion yield) was found
to have a $\nu$-dependence similar to that of the  
direct-fragmentation pions.
No correction for this process was therefore applied.

The results for the multiplicity ratio for
all charged hadrons with $z>$ 0.2
are presented as a function of $\nu$
in Fig.~\ref{fig:world} together with
data of previous experiments. In the top panel,
existing data on copper~\cite{osborne,EMC} are shown together
with the original phenomenological calculations that were
used to interpret the data. In the lower panel of
Fig.~\ref{fig:world}, the present data on $^{14}$N are
displayed together with data of previous experiments at
CERN~\cite{EMC} and SLAC~\cite{osborne} on $^{12}$C.
The SLAC data point is lower than the present data by five standard
deviations. For the SLAC data no
systematic uncertainty is available from the original
publication. Moreover, no corrections were made for the
target-mass dependence of the inclusive DIS cross section,
an effect that was not known at the time of the analysis. Such a correction
is needed because a semi-inclusive cross-section ratio was measured at
SLAC instead of the multiplicity ratio defined in Eq. (1).
An estimate of the correction for this effect -- based on the
HERMES $^{14}$N data -- results 
in a 4\% increase of $R_M^{h}$, which
reduces the discrepancy by a factor of two.

The present data for $R_M^{h}$ are observed to increase with
increasing $\nu$ and approach unity for high $\nu$, which is
consistent with the EMC data.
This behaviour suggests that for large values of $\nu$,
hadron formation takes place mainly outside the 

\begin{table} [th]
\caption{Parameters used in the phenomenological one and two
time-scale model calculations shown in Figs. 2 and 3.
In the left two columns the line type used in the figures
and the expression used for the formation time are listed, while
the three columns on the right give the values of the three
cross sections (in mb) used in the calculations.
\label{tab:pars}
}
\begin{center}
\begin{tabular}{|c|c|c|c|c|}
line type &  $t_f^h$ (fm/c) & $\sigma_q$ & $\sigma_s$ & $\sigma_h$ \\
\hline
 ---~$\cdot$~---~$\cdot$~--- & $c_h \times z\nu$ & 0 & -- & 20 \\
 ~---~---~---~---~ & $(1-\ln(z)) z\nu /(\kappa c)$ & 0.75 & 20 & 20 \\
 -~$\cdot$~-~$\cdot$~-~$\cdot$~- & $(1-\ln(z)) z\nu/(\kappa c)$ & 0 & 20 & 20 \\
 -~-~-~-~-~-~- & $(1-\ln(z)) z\nu/(\kappa c)$ & 0 & 0 & 20 \\
 $\cdot\cdot\cdot\cdot\cdot\cdot\cdot\cdot\cdot$ & $c_h \times (1-z)\nu$ & 0 & -- & 25 \\
\end{tabular}
\end{center}
\end{table}

\begin{figure} [th]
\begin{center}
\includegraphics[width=\hsize]{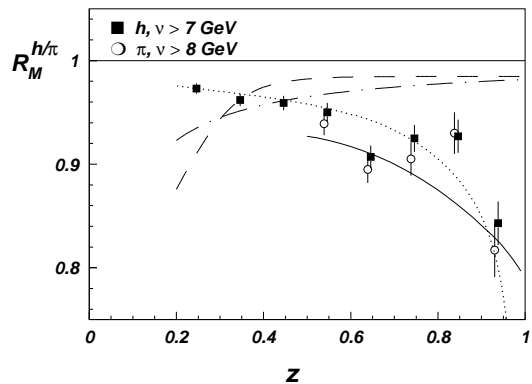}
\end{center}
\nopagebreak
\vspace{-0.5cm}
\caption[]{\label{fig:z}
The multiplicity ratio as a function of $z$ for all charged
pions (open circles) and all charged hadrons including pions (closed squares).
The full curve represents a gluon-bremsstrahlung model calculation
for pions. The dotted, dashed and dot-dashed curves represent
phenomenological formation-time calculations.}
\end{figure}

\noindent
$^{14}$N nucleus, and that the interaction of the struck quark
with the nuclear medium is weak.

For the interpretation of the data it is useful
to provide some details on the calculations from Ref.~\cite{EMC}
shown in the upper panel of Fig.~\ref{fig:world}. Similar
calculations are also used when discussing the 
$z$-dependence of the $^{14}$N data below. The long-dash dotted
curve represents a calculation in the framework of the
one time-scale model~\cite{1timescale}, in 
which it is assumed that 
the influence of the nuclear medium on the hadron multiplicity
is due to quark-nucleon scattering for $t < t_f^h$ and hadron-nucleon
scattering for $t > t_f^h$, with cross sections $\sigma_q$ and
$\sigma_h$ respectively. In the calculation shown 
$t_f^h = c_h z\nu$ with $c_h$ = 1 fm/(GeVc)~\cite{brod} has 
been assumed, and the
parameters listed in table~\ref{tab:pars} have been used.
The other curves in fig.~\ref{fig:world} are three
examples of two time-scale model calculations, in which
the formation time is assumed to be
given by $t_f^h= (1 - \ln(z)) z\nu /(\kappa c)$ with $c$ the
speed of light and $\kappa \approx$~1 GeV/fm the string 
tension~\cite{2timescale,chmaj}.
In these models an additional parameter
$\sigma_s$ is introduced that represents the interaction between
the open string and the medium occurring after 
a time interval $\tau_c = t_f^h - z\nu /(\kappa c)$. The 
expressions for $t_f^h$ and $\tau_c$ have been derived from 
the color-string model~\cite{Andersson}. The parameters used
for the three calculations shown are also listed in
table~\ref{tab:pars}.
The long-dashed curves give a fairly good
description of both the Cu and the present $^{14}$N data.
However, it should be noted that a two time-scale model
calculation with $\sigma_s = \sigma_h$ 
is equivalent to a one time-scale model calculation 
except with a different expression for $t_f^h$. 
For this reason we will use only the one time-scale
model when presenting the final parameterization of
the data near the end of this paper.

In order to further investigate the various possible
descriptions of the hadron formation process, the $z$-dependence of
$R_M^{h}$ was extracted from the present measurements for
both hadrons and identified pions.
The multiplicity ratio for hadrons (pions) with $\nu>$ 7 GeV (8 GeV)
is presented as a function of $z$ in Fig.~\ref{fig:z}. 

No significant difference between data for hadrons and pions 
is observed here. The decrease of $R_M^{h}$ with $z$, as
observed at large $z$ for the first time by the present
experiment, is at variance with the phenomenological-model
calculations, which predict an increase of $R_M^{h}$ with $z$.
This is demonstrated by the long-dash-dotted (one time-scale) and 
long-dashed (two time-scale) calculations shown in 
Fig.~\ref{fig:z}. The curves have been
obtained in the same way as the corresponding curves
in Fig.~\ref{fig:world} only using the $^{14}$N matter density
instead of the one for Cu. It is concluded that the
$(z,\nu)$-dependence of the formation time is not
given by the simple expressions~\cite{2timescale,brod} 
mentioned above. 
It is significant that the expression for $t_f^h$ implied by
the kinematic $z$-dependence of the invariant mass of the current
quark~\cite{berger} also fails to 
reproduce the observed dependence of $R_M^{h}$ on either 
$z$ or $p_T^2$~\cite{EMC,pavel}. 

The solid curve, on the other hand, gives a good account
of the data. It represents the result of a calculation for 
$R_M^{\pi}$ within the gluon-bremsstrahlung model~\cite{boris}.
Note that the calculation applies to pions only.
The basic mechanism that causes the decrease of $R_M^{h}$
at large values of 
$z$ is the following. A hadron with large 
$z$ originates from a quark that has emitted
only few gluons of relatively small energy; otherwise
its $z$-value would have been lower. As the bare quark
continuously emits gluons, the emission of only few gluons
corresponds to a small formation time. Hence,
the probability that the high-$z$ hadron is subject 

\begin{figure} [th]
\begin{center}
\includegraphics[width=\hsize]{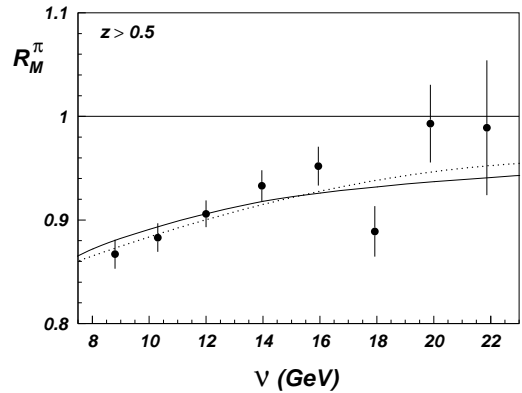}
\end{center}
\nopagebreak
\vspace{-0.5cm}
\caption[]{\label{fig:nu}
The multiplicity ratio as a function of $\nu$ for charged pions
with $z>0.5$. The solid curve represents a
gluon-bremsstrahlung model calculation. The dotted curve is the
result of a one time-scale model calculation assuming a
$(1-z) \nu$-dependence of the formation time.}
\end{figure}

\noindent
to rescattering
is largest, explaining the decrease of $R_M^h$ at high $z$.

In order to be able to study phenomenological concepts such
as the quark-nucleon cross section $\sigma_q$ and the formation
time $t_f^h$, the $z$-dependence 
of $R_M^h$ has been parameterized
in the framework of the one time-scale 
model. Using the expression
$t_f^h = c_h (1-z) \nu$, as suggested 
by the gluon-bremsstrahlung model~\cite{boris}, the dotted curve in
Fig.~\ref{fig:z} has been obtained. It represents the result of a
fit with the factor $c_h$
a free parameter, and $\sigma_q$ and $\sigma_h$ set to the
values listed in table~\ref{tab:pars}.
It is concluded that a phenomenological description of the
$z$-dependence of the $R_M^{h}$ data can be obtained if an ad-hoc
$(1-z)$-dependence of the formation time is assumed.

The $\nu$-dependence of the multiplicity ratio for charged pions 
with $z >$ 0.5 is shown in Fig.~\ref{fig:nu}. The 
calculation with the gluon-bremsstrahlung model~\cite{boris} is in fair
agreement with the data. A similar description
of the data is obtained using the 
one time-scale model with $t_f^{\pi} = c_\pi (1-z) \nu$, assuming
$\sigma_q$ = 0 mb and $\sigma_h$ = 25 mb (dotted curve in Fig.~\ref{fig:nu}).
In a two-parameter fit
with $c_\pi$ and $\sigma_q$ treated as free parameters, 
the resulting value for $\sigma_q$ is 
consistent with zero with an uncertainty of only $\pm$0.4 mb (or
$\pm$0.2 mb if the hadrons are considered).
It is concluded that in the context of the phenomenological
models, the interaction of the struck quark
with the nuclear medium is indeed very small, which is in
accordance with the results of Refs.~\cite{boris,pavel}.
Hereafter, we extract further information about the
hadron formation time within the context of the
one time-scale model with $\sigma_q \approx 0$.

\begin{figure} [th]
\begin{center}
\includegraphics[width=\hsize]{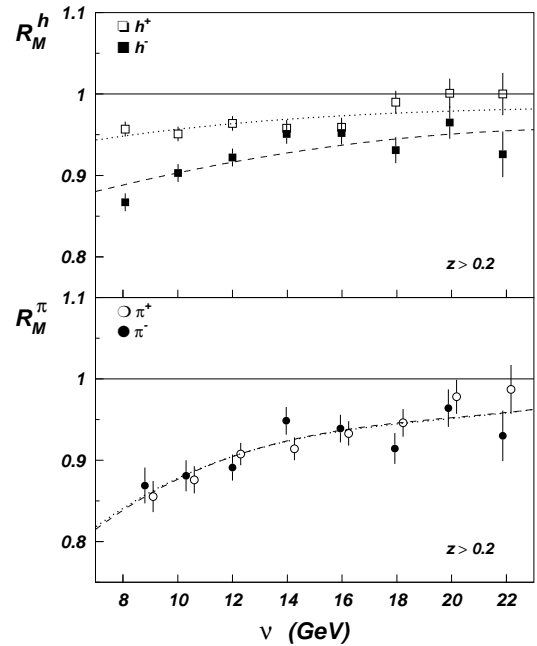}
\end{center}
\nopagebreak
\vspace{-0.5cm}
\caption[]{\label{fig:hphm}
Multiplicity ratios for hadrons including pions (top panel) and
identified pions (bottom panel) as a function of $\nu$. The
open (closed) squares represent the positive (negative)
hadrons. Identified pions are represented
by open (positive) and closed (negative) circles. The curves
are parameterizations of the data using the one time-scale
model assuming $t_f^h = c_h (1-z) \nu$.}
\end{figure}

The results presented thusfar concern the sum of positive and negative
hadrons. For both pions and hadrons the charge states can be studied 
separately. In the upper (lower) panel of Fig.~\ref{fig:hphm}, the 
multiplicity ratios for positive and negative
hadrons (pions) are displayed as a function of $\nu$. 
The same requirements on the variable $\nu$ as before
have been imposed: $\nu >$ 7 GeV for the hadron
data and $\nu >$ 8 GeV for the pion data. 
However, for the purpose of this comparison and in 
order to increase the pion statistics,
the requirement $z >$ 0.2 has been imposed in both cases.

The data show that the multiplicity ratio $R_M^\pi$ is the same for
positive and negative pions. This observation --- together with
the fact that the $\pi^{\pm}-$N interaction cross sections 
are about equal --- is consistent with
the likely assumption that the formation times for the two 
charge states are the same.

In contrast, a significant difference is observed between $R_M^{h}$ of
positive and negative hadrons. Since the multiplicity ratios
of positive and negative pions are measured to
be equal, this difference must be caused by
other hadrons, such as protons, antiprotons and kaons.
A Monte Carlo study with the LUND model~\cite{lund}
has been performed in order to estimate the relative yield
of various hadrons in the acceptance of the spectrometer.
The study shows that the contribution of antiprotons (protons)
to the negative (positive) hadron sample, averaged in the $\nu$-range 
of 3 -- 23 GeV, equals 6\% (24\%). In the same kinematic range,
the negative 
and positive kaon contributions are 10 and 
13\%, respectively.
Since the effects of the interaction between the struck
quark and the nuclear medium are found to be small, only 
two effects may account for the observed difference
between $R_M^{h+}$ and $R_M^{h-}$: the difference in the cross 
sections~\cite{PDG} for rescattering of antiprotons ($\approx$ 60~mb), 
protons ($\approx$ 40~mb), negative kaons ($\approx$ 23~mb) and 
positive kaons ($\approx$ 17~mb), or possible 
differences between 

\noindent
the formation times of baryons and mesons.
Noting that the effective hadron-nucleon cross sections, obtained
as the weighted sum of individual hadron cross sections, are practically
the same for the positive and negative hadron samples,
the observed difference of $R_M^{h^+}$ and $R_M^{h^-}$ is most
likely due to
differences in the formation times of pions,
kaons and/or (anti)protons. Since the proton 
fraction in the positive hadron sample is four times
larger than the antiproton fraction in the negative hadron sample,
it is concluded that protons have a significantly larger 
formation time than pions. 
In order to substantiate this conclusion, additional
data with good pion, kaon and proton identification are needed.

In the absence of a theoretical model predicting the observed
difference between $R_M^{h^+}$ and $R_M^{h^-}$, the data of 
Fig.~\ref{fig:hphm} have been parameterized using the 
one time-scale model~\cite{1timescale}. In these calculations,
the formation time is assumed to be given by
$t_f^h = c_h (1-z) \nu$ to ensure a proper
description of the $z$-dependence of
$R_M^{h}$, and an effective hadron-nucleon cross 
section of 25 mb is used for all hadrons.
Initially both $\sigma_q$ and $c_h$ were used as
free parameters. In all four cases (h$^+$, h$^-$, $\pi^+$,
and $\pi^-$) the fits gave a good account of the data, while
the fitted value of $\sigma_q$ was found to be
consistent with zero within uncertainties ranging from 0.3 mb
for $\pi^+$ to 1.2 mb for h$^-$. 
This implies that the dominant effect of the nucleus on
$R_M^h$ is due to hadron-nucleon interactions.
Setting $\sigma_q$ equal to zero, the fits were repeated
yielding once more a good description of the data (see Fig.~\ref{fig:hphm})
and the following set of fit parameters:
$c_{\pi^-}$ = 1.38$\pm$0.21 fm/(GeVc),
$c_{\pi^+}$ = 1.37$\pm$0.18 fm/(GeVc),
$c_{h^-}$ = 1.32$\pm$0.16 fm/(GeVc), and
$c_{h^+}$ = 3.49$\pm$0.51 fm/(GeVc).
The quoted uncertainties equal the quadratic sum of the statistical errors 
of the fit parameters and the variations of these 
parameters when the hadron-nucleon cross section is varied 
by $\pm2$ mb. The fitted
values of $c_h$ quantitatively confirm our earlier conclusions
concerning the equality of the formation times of positive and 
negative pions, and the large difference between the formation
times of positive and negative hadrons.

In summary, high accuracy data on the attenuation of charged 
hadrons and identified pions
in deep-inelastic scattering on $^{14}$N relative to $^{2}$H
have been obtained by the HERMES experiment. 
The $\nu$-dependence of the multiplicity ratio
of hadrons (h$^+$ and h$^-$) has a similar behaviour as the
CERN  muon~\cite{EMC} data.
In the previously unexplored high 
$z$-region covered by the present experiment, a
decrease of $R_M^{h}$ with $z$ is observed. This observation is 
in agreement with the
prediction of the gluon bremsstrahlung model~\cite{boris}, but is 
at variance with the prescription often
used in phenomenological models 
that the hadron formation time is proportional to
$z$, and with the $z$-dependence implied by Ref.~\cite{berger}.
In contrast to what is observed for positive and negative
pions, a difference is found in the
$\nu$-dependence of $R_M^{h}$ for positive and negative hadrons. 
This difference is interpreted to 
indicate that a proton has a larger formation time than
a pion.

Additional measurements of hadronization in various heavy nuclei with pion,
kaon and proton identification are needed to clarify the issues raised 
by the present data concerning formation-time differences of different kinds 
of hadrons. Such measurements are underway at HERMES.

We gratefully acknowledge the DESY management for its support and
the DESY staff and the staffs of the collaborating institutions.
This work was supported by the FWO-Flanders, Belgium;
the Natural Sciences and Engineering Research Council of Canada;
the INTAS and TMR network contributions from the European Community;
the German Bundesministerium f\"ur Bildung, Wissenschaft, Forschung
und Technologie; the Deutscher Akademischer Austauschdienst (DAAD);
the Italian Istituto Nazionale di Fisica Nucleare (INFN);
Monbusho International Scientific Research Program, JSPS, and Toray
Science Foundation of Japan;
the Dutch Foundation for Fundamenteel Onderzoek der Materie (FOM);
the U.K. Particle Physics and Astronomy Research Council; and
the U.S. Department of Energy and National Science Foundation.

\end{document}